\documentclass[12pt]{article}
\usepackage{amssymb}

\makeatletter
\def\numberbysection{\@addtoreset{equation}{section}
 	\def\theequation{\thesection.\arabic{equation}}}
\makeatother

\numberbysection

\newcommand{\be}{\begin{eqnarray}}
\newcommand{\ee}{\end{eqnarray}}
\newcommand{\non}{\nonumber}
\newcommand{\tr}{\mathop{\rm tr}\nolimits}
\newcommand{\id}{\mathbb{I}}
\newcommand{\csch}{\mathop{\rm csch}\nolimits}
\newcommand{\sech}{\mathop{\rm sech}\nolimits}

\newcommand{\ch}{\mathop{\rm ch}\nolimits}
\newcommand{\sh}{\mathop{\rm sh}\nolimits}
\newcommand{\tnh}{\mathop{\rm tanh}\nolimits}
\newcommand{\cth}{\mathop{\rm coth}\nolimits}

\begin{document}

\begin{titlepage}
\strut\hfill
\vspace{.5in}
\begin{center}

\LARGE Bethe ansatz of the open spin-$s$ XXZ chain\\
\LARGE with nondiagonal boundary terms\\[1.0in]
\large Rajan Murgan\footnote{e-mail: rmurgan@gustavus.edu}\\[0.8in]
\large Department of Physics, Gustavus Adolphus College\\[0.2in]  
\large St. Peter, MN 56082 USA\\

\end{center}

\vspace{.5in}

\begin{abstract}
We consider the open spin-$s$ XXZ quantum spin chain with nondiagonal boundary terms. By exploiting certain functional relations at roots of unity, we propose the Bethe ansatz solution for the transfer matrix eigenvalues for cases where atmost two of the boundary parameters are set to be arbitrary and the bulk anisotropy parameter has values $\eta = {i \pi\over 3}\,, {i \pi\over 5}\,,\ldots $. We present numerical evidence to demonstrate completeness of the Bethe ansatz solutions derived for $s = 1/2$ and $s = 1$.
\end{abstract}
\end{titlepage}

\setcounter{footnote}{0}

\section{Introduction}\label{sec:intro}

There have been significant focus of effort in solving integrable quantum spin chains for many years. In particular, integrable quantum spin chains with boundaries (integrable open quantum spin chains) have attracted much interest over the years. As a result, models such as the open XXX and XXZ quantum spin chains have been subjected to intensive studies due to their growing applications in various fields of physics, e.g. statistical mechanics, string theory and condensed matter physics. Despite numerous success in the past \cite{Gaudin}-\cite{Ma} (also refer to \cite{Do1}-\cite{Galleas} and references therein, for other related work on the subject.), there still remain unsolved problems in this area. Bethe ansatz (in its conventional form) for the most general case of the open XXZ quantum spin chain (even for the spin-1/2 case) with arbitrary nondiagonal boundary terms and generic bulk anisotropy parameter is yet to be found. In [14], Galleas found an interesting solution analogous to Bethe ansatz equations for the spin-1/2 case. This solution, written in terms of certain functional relations are expressed in terms of roots of the transfer matrix. Much progress have been made on the topic up to this point. In a series of publication, Bethe ansatz solutions have been derived for open spin-$1/2$ XXZ quantum spin chain where the boundary parameters obey certain constraint. Readers are refered to \cite{CLSW}-\cite{YZ1} for related work on the subject. Apart from this constraint, two sets of Bethe ansatz equations are needed there to obtain all $2^{N}$ eigenvalues, where $N$ is the number of sites. A special case of the above solution was generalized to open XXZ quantum spin chain with alternating spins by Doikou \cite{Do2} using the functional relation approach, proposed by Nepomechie in \cite{Ne} to solve the spin-$1/2$ case (which indeed the method used in this paper). In \cite{Do3}, related work was carried out using the method in \cite{CLSW}. Recently in \cite{FNR}, Frappat et al. further generalized the spin-$1/2$ XXZ Bethe ansatz solution (for boundary parameters obeying the constraint) to the spin-$s$ case by utilizing an approach based on $Q$-operator and $T$-$Q$ equation, which was developed earlier for the spin-$1/2$ XXZ chain in \cite{YNZ} and subsequently applied to the spin-$1/2$ XYZ chain in \cite{YZ3}. As in the spin-$1/2$ case, two sets of Bethe ansatz equations are also needed there to produce all $(2s+1)^{N}$ eigenvalues, where again $N$ represents the number of sites.

In this paper, we present Bethe ansatz solutions for open spin-$s$ XXZ quantum spin chain without such a constraint among the boundary parameters. We follow similar approach as given in \cite{Ne, NR, MNS} that was used to solve the $s = 1/2$ case. It is based on fusion \cite{MNR, fusion, fusion2}, the truncation of the fusion hierarchy at roots of unity \cite{truncation} and the Bazhanov-Reshetikin \cite{BR} solution of the RSOS models. As in \cite{MNS}, there are atmost two arbitrary boundary parameters. The rest of the parameters are fixed to some values. The approach we use, which is based on functional relations obeyed by transfer matrix at roots of unity \cite{Ne} yields Bethe ansatz solution which gives completely all the $(2s + 1)^{N}$ eigenvalues. One limitation of the solution is that it is valid only at roots of unity, namely when the bulk anisotropy parameter has values $\eta = {i \pi\over p+1}$. In this paper, we consider only even values of $p$. Lack of single set of Bethe ansatz equations that yield complete eigenvalues for the model considered here, namely where the boundary parameters are arbitrary (even at most two) has motivated us to study this problem. Moreover, we note that the relation of $s = 1$ case to the supersymmetric sine-Gordon (SSG) model \cite{SSG} (here the boundary version \cite{BSSG, ANS}), has also been part of our motivation for considering the problem.   

The outline of the paper is as follows: In Sec. 2, we review the construction of the so-called fused $R$ \cite{fusion, Ka, spinsXXX, spinsXXZ} and $K^{\mp}$ \cite{MNR, fusion2} matrices from the corresponding spin-$1/2$ matrices. For some original work on spin-$1/2$ $K^{\mp}$ matrices, refer to \cite{dVGR, GZ}. Construction of commuting transfer matrices from these fused matrices (using Sklyanin's work \cite{Sk}, which in turn relies on Cherednik's previous results \cite{Ch}), together with some of their properties are reviewed. Fusion hierachy and functional relations obeyed by transfer matrices are also reviewed. In Sec. 3, we present the Bethe ansatz solutions for cases with atmost two arbitrary boundary parameters at roots of unity, e.g. $\eta = {i \pi\over 3}\,, {i\pi\over 5}\,,\ldots$, by exploiting the reviewed functional relations obeyed by the transfer matrices. Further, we present numerical results in Sec. 4 to illustrate the completeness of our solution, using $s = 1/2$ and $s = 1$ as examples, where the Bethe roots and energy eigenvalues derived from the Bethe ansatz equations (for some values of $p$ and $N$) are given. We remark that these energy eigenvalues coincide with the ones obtained from direct diagonalization of the Hamiltonians. Finally, we conclude the paper with discussion of the results and potential future works in Sec. 5.

\section{Transfer matrices, fusion hierachy and functional relations at roots of unity}\label{sec:transfer}

In this section, in order to make the paper relatively self-contained, we review some crucial concepts on the construction of commuting transfer matrices for $N$-site open spin-$s$ XXZ quantum spin chain. Materials reviewed here on fused $R$, $K^{\mp}$ and higher spin transfer matrices are borrowed from \cite{FNR}, as presented there. As constructed in \cite{Sk}, the commuting transfer matrix for $s=1/2$, which we denote (following notations adopted in \cite{FNR}) by
$t^{(\frac{1}{2},\frac{1}{2})}(u)$, whose auxiliary space as well as each of
its $N$ quantum spaces are two-dimensional, one can similarly construct a transfer matrix $t^{(j,s)}(u)$ whose
auxiliary space is spin-$j$ ($(2j+1)$-dimensional) and each of its $N$
quantum spaces are spin-$s$ ($(2s+1)$-dimensional), for any $j,s \in 
\{\frac{1}{2},1,\frac{3}{2},\ldots\}$ using the so-called fused $R$ \cite{fusion, Ka, spinsXXX, spinsXXZ} and $K^{\mp}$ \cite{MNR, fusion2} matrices. As for the spin-$1/2$ case, these $R$ and $K^{\mp}$ \ matrices serve as building blocks in the construction of the commuting transfer matrices for higher spins. We list them below along with some of their properties. The fused-$R$ matrices can be constructed as given below, 
\be
R^{(j,s)}_{\{a\} \{b\}}(u) =
P_{\{a\}}^{+} P_{\{b\}}^{+} 
\prod_{k=1}^{2j}\prod_{l=1}^{2s}
R^{(\frac{1}{2},\frac{1}{2})}_{a_{k} b_{l}}(u+(k+l-j-s-1)\eta)\, 
P_{\{a\}}^{+} P_{\{b\}}^{+} \,,
\label{fusedRmatrix}
\ee 
where $\{a\} = \{a_{1}, \ldots , a_{2j}\}$, $\{b\} = \{b_{1}, \ldots , 
b_{2s}\}$, and $P_{\{a\}}^{+}$ is the symmetric projector given by 
\be
P_{\{a\}}^{+} ={1\over (2j)!} 
\prod_{k=1}^{2j}\left(\sum_{l=1}^{k}{\cal P}_{a_{l}, a_{k}} \right) \,,
\label{projector}
\ee
${\cal P}$ is the permutation operator, with ${\cal P}_{a_{k}, 
a_{k}} \equiv 1$; Similar definition also holds for $P_{\{b\}}^{+}$. 
$R^{(\frac{1}{2},\frac{1}{2})}(u)$ is given by
\be
R^{(\frac{1}{2},\frac{1}{2})}(u) = \left( \begin{array}{cccc}
	\sh  (u + \eta) &0            &0           &0            \\
	0                 &\sh  u     &\sh \eta  &0            \\
	0                 &\sh \eta   &\sh  u    &0            \\
	0                 &0            &0           &\sh  (u + \eta)
\end{array} \right) \,,
\label{Rmatrix}
\ee 
where $\eta$ is the bulk anisotropy parameter. Note that the fundamental $R$ matrix 
satisfies the following unitarity relation 
\be
R^{(\frac{1}{2},\frac{1}{2})}(u) R^{(\frac{1}{2},\frac{1}{2})}(-u) = 
- \xi(u) 1 \,, \qquad \xi(u) = \sh(u+ \eta) \sh(u - \eta) \,.
\label{xi}
\ee
The $R$ matrices in the product (\ref{fusedRmatrix}) are ordered 
in the order of increasing $k$ and $l$.
The fused $R$ matrices satisfy the Yang-Baxter equations \cite{Yang1}
\be
R^{(j,k)}_{\{a\} \{b\}}(u-v)\, R^{(j,s)}_{\{a\} \{c\}}(u)\, 
R^{(k,s)}_{\{b\} \{c\}}(v) =
R^{(k,s)}_{\{b\} \{c\}}(v)\,  R^{(j,s)}_{\{a\} \{c\}}(u)\, 
R^{(j,k)}_{\{a\} \{b\}}(u-v) \,.
\ee 
Having defined fused-R matrices, one can analogously construct fused $K^{-}$ matrices \cite{MNR, fusion2}
\be
K^{- (j)}_{\{a\}}(u) &=& P_{\{a\}}^{+} \prod_{k=1}^{2j} \Bigg\{ \left[ 
\prod_{l=1}^{k-1} R^{(\frac{1}{2},\frac{1}{2})}_{a_{l}a_{k}}
(2u+(k+l-2j-1)\eta) \right] \non \\
&\times & K^{- (\frac{1}{2})}_{a_{k}}(u+(k-j-\frac{1}{2})\eta) \Bigg\}
P_{\{a\}}^{+} \,,
\label{fusedKmatrix}
\ee 
where $K^{- (\frac{1}{2})}(u)$ is the
$2 \times 2$ matrix whose components
are given by \cite{dVGR, GZ}
\be
K_{11}^{-}(u) &=& 2 \left( \sh \alpha_{-} \ch \beta_{-} \ch u +
\ch \alpha_{-} \sh \beta_{-} \sh u \right) \non \\
K_{22}^{-}(u) &=& 2 \left( \sh \alpha_{-} \ch \beta_{-} \ch u -
\ch \alpha_{-} \sh \beta_{-} \sh u \right) \non \\
K_{12}^{-}(u) &=& e^{\theta_{-}} \sh  2u \,, \qquad 
K_{21}^{-}(u) = e^{-\theta_{-}} \sh  2u \,,
\label{Kminuscomponents}
\ee
where $\alpha_{-} \,, \beta_{-} \,, \theta_{-}$ are the boundary
parameters.
The products of braces $\{ \ldots \}$ in (\ref{fusedKmatrix})
are ordered in the order of increasing $k$.  
The fused $K^{-}$ matrices satisfy the boundary Yang-Baxter equations 
\cite{Ch}
\be
\lefteqn{R^{(j,s)}_{\{a\} \{b\}}(u-v)\, K^{- (j)}_{\{a\}}(u)\,
R^{(j,s)}_{\{a\} \{b\}}(u+v)\, K^{- (j)}_{\{b\}}(v)}\non \\
& & =K^{- (j)}_{\{b\}}(v)\, R^{(j,s)}_{\{a\} \{b\}}(u+v)\,
K^{- (j)}_{\{a\}}(u)\, R^{(j,s)}_{\{a\} \{b\}}(u-v) \,.
\ee
The fused $K^{+}$ matrices are given by
\be
K^{+ (j)}_{\{a\}}(u)  = {1\over f^{(j)}(u)}\,K^{- (j)}_{\{a\}}
(-u-\eta)\Big\vert_{(\alpha_-,\beta_-,\theta_-)\rightarrow
(-\alpha_+,-\beta_+,\theta_+)} \,,
\ee
where the normalization factor is, 
\be
f^{(j)}(u) = \prod_{l=1}^{2j-1}\prod_{k=1}^{l}
[-\xi( 2u + (l+k+1-2j)\eta) ] 
\label{Kplusnormalization}
\ee
Using the above results, one can construct the transfer matrix $t^{(j,s)}(u)$,
\be
t^{(j,s)}(u) = \tr_{\{a\}} K^{+ (j)}_{\{a\}}(u)\,
T^{(j,s)}_{\{a\}}(u)\, K^{- (j)}_{\{a\}}(u)\,
\hat T^{(j,s)}_{\{a\}}(u) \,,
\ee 
where the monodromy matrices are given by products of $N$ fused $R$ 
matrices, 
\be
T^{(j,s)}_{\{a\}}(u) &=& R^{(j,s)}_{\{a\}, \{b^{[N]}\}}(u) \ldots 
R^{(j,s)}_{\{a\}, \{b^{[1]}\}}(u) \,, \non \\
\hat T^{(j,s)}_{\{a\}}(u) &=& R^{(j,s)}_{\{a\}, \{b^{[1]}\}}(u) \ldots
R^{(j,s)}_{\{a\}, \{b^{[N]}\}}(u) \,.
\ee 
These transfer matrices commute for different
values of spectral parameter for any $j , j' \in \{\frac{1}{2}, 1,
\frac{3}{2}, \ldots \}$ and any $s \in \{\frac{1}{2}, 1, \frac{3}{2},
\ldots \}$,
\be
\left[ t^{(j,s)}(u) \,, t^{(j',s)}(u') \right] = 0 \,.
\label{commutativity}
\ee
Furthermore, they also obey the fusion hierarchy \cite{MNR, fusion2, FNR}\footnote{See the appendix in \cite{FNR} for more details on the fusion hierachy.}  
\be
t^{(j-\frac{1}{2},s)}(u- j\eta)\, t^{(\frac{1}{2},s)}(u) =
t^{(j,s)}(u-(j-\frac{1}{2})\eta)  + \delta^{(s)}(u-\eta)\,
t^{(j-1,s)}(u-(j+\frac{1}{2})\eta) \,, 
\label{hierarchy}
\ee
$j = 1,\frac{3}{2},\ldots$, where $t^{(0,s)}=1$, and
$\delta^{(s)}(u)$ is given by
\be
\delta^{(s)}(u) &=& 
\delta_{0}^{(s)}(u)\delta_{1}^{(s)}(u)
\label{dd}
\ee
where 
\be
\delta_{0}^{(s)}(u) &=& \left[\prod_{k=0}^{2s-1}\xi(u+(s-k+\frac{1}{2})\eta)\right]^{2N} 
{\sh(2u) \sh(2u+4\eta)\over \sh(2u+\eta) \sh(2u+3\eta)}\non \\
\delta_{1}^{(s)}(u) &=& 2^{4}\sh(u+\alpha_{-}+\eta)\sh(u-\alpha_{-}+\eta)\ch(u+\beta_{-}+\eta)\ch(u-\beta_{-}+\eta)\non \\
&\times& \sh(u+\alpha_{+}+\eta)\sh(u-\alpha_{+}+\eta)\ch(u+\beta_{+}+\eta)\ch(u-\beta_{+}+\eta) \,.
\label{delta01}
\ee
Note that the $\delta^{(s)}(u)$ in \cite{FNR} differs to the one given here merely by a shift in $\eta$.

Next, we list few important
properties of the rescaled ``fundamental'' transfer matrix
$\tilde t^{(\frac{1}{2},s)}(u)$ (defined below), which are useful in determining
its eigenvalues. Following the definition of $\tilde t^{(\frac{1}{2},s)}(u)$ as in \cite{FNR}, we have
\be
\tilde t^{(\frac{1}{2},s)}(u) = {1\over g^{(\frac{1}{2},s)}(u)^{2N}} 
t^{(\frac{1}{2},s)}(u) \,,
\label{tildet}
\ee
where
\be
g^{(\frac{1}{2},s)}(u) = \prod_{k=1}^{2s-1} \sh(u+(s-k+\frac{1}{2})\eta)
\label{gfunction}
\ee
This transfer matrix has following useful properties: 
\be
\tilde t^{(\frac{1}{2},s)}(u + i\pi) = \tilde 
t^{(\frac{1}{2},s)}(u) \qquad (i\pi\mbox{ - periodicity}) 
\label{periodicity}
\ee
\be 
\tilde t^{(\frac{1}{2},s)}(-u -\eta) = \tilde 
t^{(\frac{1}{2},s)}(u) \qquad (\mbox{crossing}) 
\label{crossing}
\ee
\be
\tilde t^{(\frac{1}{2},s)}(0) = -2^{3}\sh^{2N}((s+\frac{1}{2})\eta) 
\ch \eta \sh \alpha_{-} \ch \beta_{-} \sh \alpha_{+} \ch \beta_{+} \id
\quad (\mbox{initial condition} )
\label{initial}
\ee
\be
\tilde t^{(\frac{1}{2},s)}(u)\Big\vert_{\eta=0} &=& 
2^{3}\sh^{2N}u \Big[ -\sh \alpha_{-} \ch \beta_{-} \sh \alpha_{+} \ch 
\beta_{+} \ch^{2}u  \non \\
&+& \ch \alpha_{-} \sh \beta_{-} \ch \alpha_{+} \sh 
\beta_{+} \sh^{2}u \non \\
&-& \ch(\theta_{-}-\theta_{+}) \sh^{2}u \ch^{2}u 
\Big] \id \quad (\mbox{semi-classical limit} )
\label{semiclassical}
\ee
\be 
\tilde t^{(\frac{1}{2},s)}(u) &\sim& -{1\over 2^{2N+1}}e^{(2N+4)u 
+(N+2)\eta} \ch(\theta_{-}-\theta_{+}) \id \quad \mbox{for} \ u\rightarrow 
+\infty \non \\
& & \qquad \qquad \qquad \qquad \qquad (\mbox{asymptotic behavior} ) 
\label{asymptotic}
\ee
where $\id$ is the identity matrix.

Due to the commutativity property (\ref{commutativity}), the corresponding simultaneous eigenvectors are independent of the spectral 
parameter. Hence, (\ref{periodicity}) - (\ref{asymptotic}) hold for the corresponding eigenvalues as well. In addition to the above mentioned properties, for bulk anisotropy values $\eta = {i \pi\over p+1}$, with $p= 1 \,, 2 \,, \ldots $, the ``fundamental'' transfer matrix, $t^{(\frac{1}{2},s)}(u)$ (and hence each of the corresponding eigenvalues, $\Lambda^{(\frac{1}{2},s)}(u)$) obeys functional relations of order $p+1$ \cite{Ne}
\be
\lefteqn{t^{(\frac{1}{2},s)}(u) t^{(\frac{1}{2},s)}(u +\eta) \ldots t^{(\frac{1}{2},s)}(u + p \eta)} \non \\
&-& \delta^{(s)} (u-\eta) t^{(\frac{1}{2},s)}(u +\eta) t^{(\frac{1}{2},s)}(u +2\eta) 
\ldots t^{(\frac{1}{2},s)}(u + (p-1)\eta) \non \\
&-& \delta^{(s)} (u) t^{(\frac{1}{2},s)}(u +2\eta) t^{(\frac{1}{2},s)}(u +3\eta)
\ldots t^{(\frac{1}{2},s)}(u + p \eta) \non \\
&-& \delta^{(s)} (u+\eta) t^{(\frac{1}{2},s)}(u) t^{(\frac{1}{2},s)}(u +3\eta) t^{(\frac{1}{2},s)}(u +4\eta) 
\ldots t^{(\frac{1}{2},s)}(u + p \eta) \non \\
&-& \delta^{(s)} (u+2\eta) t^{(\frac{1}{2},s)}(u) t^{(\frac{1}{2},s)}(u +\eta) t^{(\frac{1}{2},s)}(u +4\eta) 
\ldots t^{(\frac{1}{2},s)}(u + p \eta) - \ldots \non \\
&-& \delta^{(s)} (u+(p-1)\eta) t^{(\frac{1}{2},s)}(u) t^{(\frac{1}{2},s)}(u +\eta) 
\ldots t^{(\frac{1}{2},s)}(u +  (p-2)\eta) \non \\
&+& \ldots  = f(u) \,.
\label{funcrltn}
\ee 
For example, for $p=2$ and $p=4$, the functional relations are
\be
& & t^{(\frac{1}{2},s)}(u) t^{(\frac{1}{2},s)}(u+\eta) t^{(\frac{1}{2},s)}(u+2\eta)
- \delta^{(s)}(u-\eta) t^{(\frac{1}{2},s)}(u+\eta) 
- \delta^{(s)}(u) t^{(\frac{1}{2},s)}(u+2\eta)\non \\
& &\quad - \delta^{(s)}(u+\eta) t^{(\frac{1}{2},s)}(u)
= f(u) \,.
\label{funcrltn2}
\ee
and
\be 
& & t^{(\frac{1}{2},s)}(u) t^{(\frac{1}{2},s)}(u+\eta) t^{(\frac{1}{2},s)}(u+2\eta)t^{(\frac{1}{2},s)}(u+3\eta) t^{(\frac{1}{2},s)}(u+4\eta)\non \\  
& &\quad + \delta^{(s)}(u+\eta) \delta^{(s)}(u-2\eta) t^{(\frac{1}{2},s)}(u) 
+ \delta^{(s)}(u) \delta^{(s)}(u+2\eta) t^{(\frac{1}{2},s)}(u+4\eta)\non \\ 
& &\quad + \delta^{(s)}(u+\eta) \delta^{(s)}(u-\eta) t^{(\frac{1}{2},s)}(u+3\eta) 
- \delta^{(s)}(u+\eta) t^{(\frac{1}{2},s)}(u) t^{(\frac{1}{2},s)}(u+3\eta) t^{(\frac{1}{2},s)}(u+4\eta)\non \\  
& &\quad + \delta^{(s)}(u) \delta^{(s)}(u-2\eta) t^{(\frac{1}{2},s)}(u+2\eta)
- \delta^{(s)}(u) t^{(\frac{1}{2},s)}(u+2\eta) t^{(\frac{1}{2},s)}(u+3\eta) t^{(\frac{1}{2},s)}(u+4\eta)\non \\
& &\quad + \delta^{(s)}(u-\eta) \delta^{(s)}(u+2\eta) t^{(\frac{1}{2},s)}(u+\eta)
- \delta^{(s)}(u+2\eta) t^{(\frac{1}{2},s)}(u) t^{(\frac{1}{2},s)}(u+\eta) t^{(\frac{1}{2},s)}(u+4\eta)\non \\
& &\quad - \delta^{(s)}(u-2\eta) t^{(\frac{1}{2},s)}(u) t^{(\frac{1}{2},s)}(u+\eta) t^{(\frac{1}{2},s)}(u+2\eta)\non \\
& &\quad - \delta^{(s)}(u-\eta) t^{(\frac{1}{2},s)}(u+\eta) t^{(\frac{1}{2},s)}(u+2\eta) t^{(\frac{1}{2},s)}(u+3\eta)
= f(u) \,.
\label{funcrltn4}
\ee
respectively. The scalar function $f(u)$ (which can be expressed as $f(u) = f_{0}(u)f_{1}(u)$) is given in terms of the boundary parameters $\alpha_{\mp} \,, \beta_{\mp} \,, \theta_{\mp}$ (for even $p$) by  
\be
f_{0}(u)  = \left\{ 
\begin{array}{ll}
    (-1)^{N+1} 2^{-4 s p N}\sh^{4sN} \left( (p+1)u \right)\,, \\
    \qquad \qquad s= {1\over 2}\,, {3\over 2}\,, {5\over 2}\,, \ldots \\
   (-1)^{N+1} 2^{-4 s p N} \ch^{4sN} \left( (p+1)u \right)\,, \\
\qquad \qquad s= 1\,, 2\,, 3\,, \ldots \\
\end{array} \right.
\label{f0}
\ee
and
\be
f_{1}(u) &=& (-1)^{N+1} 2^{3-2 p} \Big( \non \\
& & \hspace{-0.2in}
\sh \left( (p+1) \alpha_{-} \right)\ch \left( (p+1) \beta_{-} \right)
\sh \left( (p+1) \alpha_{+} \right)\ch \left( (p+1) \beta_{+} \right)
\ch^{2} \left( (p+1)u \right) \non \\
&-&
\ch \left( (p+1) \alpha_{-} \right)\sh \left( (p+1) \beta_{-} \right)
\ch \left( (p+1) \alpha_{+} \right)\sh \left( (p+1) \beta_{+} \right)
\sh^{2} \left( (p+1)u \right) \non \\
&-&
(-1)^{N} \ch \left( (p+1)(\theta_{-}-\theta_{+}) \right)
\sh^{2} \left( (p+1)u \right) \ch^{2} \left( (p+1)u \right) 
\Big) \,.
\label{f1}
\ee
for $s = {1\over 2}\,, {3\over 2}\,, {5\over 2}\ldots$ 
and 
\be
f_{1}(u) &=& (-1)^{N+1} 2^{3-2 p} \Big( \non \\
& & \hspace{-0.2in}
\sh \left( (p+1) \alpha_{-} \right)\ch \left( (p+1) \beta_{-} \right)
\sh \left( (p+1) \alpha_{+} \right)\ch \left( (p+1) \beta_{+} \right)
\ch^{2} \left( (p+1)u \right) \non \\
&-&
\ch \left( (p+1) \alpha_{-} \right)\sh \left( (p+1) \beta_{-} \right)
\ch \left( (p+1) \alpha_{+} \right)\sh \left( (p+1) \beta_{+} \right)
\sh^{2} \left( (p+1)u \right) \non \\
&-&
 \ch \left( (p+1)(\theta_{-}-\theta_{+}) \right)
\sh^{2} \left( (p+1)u \right) \ch^{2} \left( (p+1)u \right) 
\Big) \,.
\label{f1ints}
\ee
for $s = 1\,, 2\,, 3\ldots$,
Note that $f(u)$ satisfies
\be
f(u + \eta) = f(u) \,, \qquad f(-u)=f(u) \,.
\ee
and
\be
f_{0}(u)^{2} = \prod_{j=0}^{p}\delta^{(s)}_{0}(u + j \eta) \,.
\label{identity}
\ee
where $\delta^{(s)}_{0}(u)$ is given by (\ref{delta01}).

\section{Bethe ansatz}

In this section, we give main results of this paper. We derive Bethe ansatz equations for various cases where atmost two of the boundary parameters $\{ \alpha_{-}, \alpha_{+},
\beta_{-}, \beta_{+} \}$ are arbitrary by adopting the steps given in \cite{MNS}. By considering atmost two boundary parameters, we find certain factors in the calculation become perfect squares. This facilitate the computations that follow. More on this is explained below.

\subsection{$\alpha_{+}$, $\alpha_{-}$ arbitrary}\label{subsec:alphapm}

Here, we take both $\alpha_{-}$ and $\alpha_{+}$ to be arbitrary while setting $\beta_{\pm} = \eta$, $\theta_{-} = \theta_{+} = \theta$, where $\theta$ is arbitrary. In order to obtain Bethe ansatz equations for the transfer matrix eigenvalues $\Lambda^{(\frac{1}{2},s)}(u)$, we shall recast the functional relations (\ref{funcrltn}) as the condition that the determinant of a certain matrix vanishes (following \cite{BR}). We find that the functional relations (\ref{funcrltn}) for the transfer matrix
eigenvalues can be written as
\be
\det {\cal M} = 0 \,,
\label{det}
\ee
where ${\cal M}$ is given by the $(p+1) \times (p+1)$ matrix
\be
{\cal M} = \left(
\begin{array}{cccccccc}
    \Lambda^{(\frac{1}{2},s)}(u) & -h(u) & 0  & \ldots  & 0 & -h(-u+p \eta)  \\
    -h(-u) & \Lambda^{(\frac{1}{2},s)}(u+p\eta) & -h(u+p \eta)  & \ldots  & 0 & 0  \\
    \vdots  & \vdots & \vdots & \ddots 
    & \vdots  & \vdots    \\
   -h(u+p^{2} \eta)  & 0 & 0 & \ldots  & -h(-u-p(p-1) \eta) &
    \Lambda^{(\frac{1}{2},s)}(u+p^{2}\eta)
\end{array} \right)  \,,
\label{calMalpha}
\ee
(whose successive rows are obtained by simultaneously shifting $u \mapsto u+ p \eta$
and cyclically permuting the columns to the right)
provided that there exists a function $h(u)$ with the following properties
\be
h(u + 2 i \pi) = h \left(u +2(p+1)\eta \right) &=& h(u) \,, \label{cond0} \\
h(u+(p+2)\eta)\ h(-u-(p+2)\eta) &=& \delta^{(s)}(u) \,, \label{cond1} \\
\prod_{j=0}^{p} h(u+2j\eta) + \prod_{j=0}^{p} h(-u-2j\eta) &=& f(u) 
\,. \label{cond2} 
\ee
From (\ref{cond0})-(\ref{cond2}), we see that the problem of finding $h(u)$ then reduces to solving the following quadratic equation in $z(u)$,
\be
z(u)^{2} -z(u) f(u) +  \prod_{j=0}^{p} \delta^{(s)}\left(u+(2j-1)\eta\right) = 0 \,,
\label{quadratic}
\ee
where 
\be
z(u) = \prod_{j=0}^{p} h(u+2j\eta) \,.
\label{producth1}
\ee
For the cases considered here and in subsequent sections, the discriminants of the corresponding
quadratic equations are perfect squares, and the factorizations such as (\ref{producth1}) can be readily carried out. However, when all boundary parameters are arbitrary, the discriminant is no longer a perfect square; and factoring the result becomes a formidable challenge. Solving the quadratic equation (\ref{quadratic}) for $z(u)$, making use of the
explicit expressions (\ref{delta01}) and (\ref{f0})-(\ref{f1ints}) for $\delta^{(s)}(u)$
and $f(u)$, respectively, we obtain the following for $h(u)$,

\be
h(u) = h_{0}(u) h_{1}(u) \,,
\label{h}
\ee
with 
\be
h_{0}(u) = (-1)^{2sN} 4\left[\prod_{k=0}^{2s-1}\sh(u+(s-k+\frac{1}{2})\eta)\right]^{2N} 
{\sh(2u+2\eta)\over \sh(2u+\eta)}
\label{h0alphamp}
\ee
and
\be
h_1(u) = \left\{ 
\begin{array}{ll}
    \ch^{2}(u-\eta)\sh(u-\alpha_{-})\sh(u+(-1)^{N}\alpha_{+}){\ch\left({1\over 2}(u+\alpha_{-}+\eta) \right)\over 
 \ch\left({1\over 2}(u-\alpha_{-}-\eta) \right)}{\ch\left({1\over 2}(u+(-1)^{N+1}\alpha_{+}+\eta) \right)\over 
 \ch\left({1\over 2}(u+(-1)^{N}\alpha_{+}-\eta) \right)}\,, \\
    \qquad \qquad s= {1\over 2}\,, {3\over 2}\,, {5\over 2}\,, \ldots \\
   \ch^{2}(u-\eta)\sh(u-\alpha_{-})\sh(u+\alpha_{+}){\ch\left({1\over 2}(u+\alpha_{-}+\eta) \right)\over 
 \ch\left({1\over 2}(u-\alpha_{-}-\eta) \right)}{\ch\left({1\over 2}(u-\alpha_{+}+\eta) \right)\over 
 \ch\left({1\over 2}(u+\alpha_{+}-\eta) \right)}\,, \\
\qquad \qquad s= 1\,, 2\,, 3\,, \ldots \\
\end{array} \right.
\label{halpha}
\ee
Further, the structure of the matrix ${\cal M}$ (\ref{calMalpha}) suggests that its 
null eigenvector has the form $\big( Q(u)\,, Q(u+p\eta) \,, \ldots
\,, Q(u+p^{2}\eta) \big)$, where $Q(u)$ has the periodicity property
\be
Q(u + 2i\pi) = Q(u) \,.
\label{Qperiodicity}
\ee
It suggests that the transfer matrix
eigenvalues are given by
\be
\Lambda^{(\frac{1}{2},s)}(u) = h(u) {Q(u + p\eta)\over Q(u)} 
+ h(-u+p \eta) {Q(u -p\eta)\over Q(u)}  \,,
\label{eigenvalues} 
\ee
which is of the Baxter's $TQ$ relation form.
Noting that the functions $h(u)$
and $h(-u+p\eta)$ (see (\ref{h})-(\ref{halpha})) have the factor
$g^{(\frac{1}{2},s)}(u)^{2N}$ (see (\ref{gfunction})) in common (since $g^{(\frac{1}{2},s)}(u)=g^{(\frac{1}{2},s)}(-u+p\eta)$), we can rewrite (\ref{eigenvalues}) in terms of  the eigenvalues of $\tilde t^{(\frac{1}{2},s)}(u)$ (see (\ref{tildet})) as
\be
\tilde \Lambda^{(\frac{1}{2},s)}(u) = 
\tilde h(u) {Q(u+p\eta)\over Q(u)} +
\tilde h(-u+p \eta) {Q(u -p\eta)\over Q(u)} \,,
\label{TQ3}
\ee
where
\be 
\tilde h(u) &=&  \tilde h_{0}(u) h_{1}(u)
\label{haa}
\ee
with
\be
\tilde h_{0}(u) = (-1)^{2sN} 4\sh^{2N}(u+(s+\frac{1}{2})\eta) 
{\sh(2u+2\eta)\over \sh(2u+\eta)}
\label{h0gen}
\ee
and
\be
Q(u) = \prod_{j=1}^{M} 
\sh \left( {1\over 2}(u - u_{j}) \right)
\sh \left( {1\over 2}(u + u_{j} - p\eta) \right) \,,
\label{Q}
\ee 
with the periodicity (\ref{Qperiodicity}) as well as the crossing
property 
\be
Q(-u+ p\eta) = Q(u) \,.
\label{Qcrossing}
\ee
where
\be
M=2sN+2p+1
\,, \label{M}
\ee
which is confirmed numerically for small values of $N$ and $p$. We stress here that the $h(u)$ given above is not the only solution. It is obtained largely by trial and error, verifying numerically for small values of $N$ that the eigenvalues can indeed be expressed as (\ref{TQ3}) with $Q(u)$'s of the form given by (\ref{Q}). We also remark that (\ref{M}) is consistent with the asymptotic behavior (\ref{asymptotic}). Making use of the analyticity of $\tilde \Lambda^{(\frac{1}{2},s)}(u)$, we have the following for the Bethe ansatz equations,
\be
{\tilde h(u_{j})\over \tilde h(-u_{j}+p\eta)} = 
-{Q(u_{j}-p\eta)\over Q(u_{j}+p\eta)} \,, 
\qquad j = 1 \,, \ldots \,, M \,.
\label{BAeqs}
\ee

\subsection{$\beta_{+}$, $\beta_{-}$ arbitrary}\label{subsec:betapm}

In the following, we set $\beta_{-}$ and $\beta_{+}$ arbitrary while setting $\alpha_{\pm} = \eta$, $\theta_{-} = \theta_{+} = \theta$. As before, we write the functional relations (\ref{funcrltn}) for the transfer matrix eigenvalues in the form of (\ref{det}), where for this case, the matrix ${\cal M}$ is given by

${\cal M} =$
\be
\left(
\begin{array}{cccccccc}
    \Lambda^{(\frac{1}{2},s)}(u) & -h(u) & 0  & \ldots  & 0 & -h(-u- \eta)  \\
    -h(-u-(p+1)\eta) & \Lambda^{(\frac{1}{2},s)}(u+p\eta) & -h(u+p \eta)  & \ldots  & 0 & 0  \\
    \vdots  & \vdots & \vdots & \ddots 
    & \vdots  & \vdots    \\
   -h(u+p^{2} \eta)  & 0 & 0 & \ldots  & -h(-u-(p^{2}+1) \eta) &
    \Lambda^{(\frac{1}{2},s)}(u+p^{2}\eta)
\end{array} \right)  \,,
\label{calMbeta}\non \\
\ee
if $h(u)$ satisfies
\be
h(u + 2 i \pi) = h \left(u +2(p+1)\eta \right) &=& h(u) \,,
\label{cond0beta} \\
h(u+(p+2)\eta)\ h(-u-\eta) &=& \delta^{(s)}(u) \,, \label{cond1beta} \\
\prod_{j=0}^{p} h(u+2j\eta) + \prod_{j=0}^{p} h(-u-(2j+1)\eta) &=& f(u) 
\,. \label{cond2beta} 
\ee
Proceeding in a similar way to the previous case and setting $h(u) = h_{0}(u) h_{1}(u)$ we find 
\be
h_{0}(u) = (-1)^{2sN} 4\left[\prod_{k=0}^{2s-1}\sh(u+(s-k+\frac{1}{2})\eta)\right]^{2N} 
{\sh(2u+2\eta)\over \sh(2u+\eta)}
\label{h0betamp}
\ee
and
\be
h_1(u) = \left\{ 
\begin{array}{ll}
    \sh(u-\eta)\sh(u+\eta)(\ch u-i\sh\beta_{-})(\ch u+(-1)^{N}i\sh\beta_{+})\,, \\
    \qquad \qquad s= {1\over 2}\,, {3\over 2}\,, {5\over 2}\,, \ldots \\
   \sh(u-\eta)\sh(u+\eta)(\ch u+i\sh\beta_{-})(\ch u-i\sh\beta_{+})\}\,, \\
\qquad \qquad s= 1\,, 2\,, 3\,, \ldots \\
\end{array} \right.
\label{hbeta}
\ee
The transfer matrix eigenvalues are now given by
\be
\Lambda^{(\frac{1}{2},s)}(u) = h(u) {Q(u + p\eta)\over Q(u)} 
+ h(-u - \eta) {Q(u -p\eta)\over Q(u)}  \,,
\label{eigenvaluesbeta} 
\ee
As before, due to the common factor $g^{(\frac{1}{2},s)}(u)^{2N}$ (see \ref{gfunction}), and using the crossing symmetry $g^{(\frac{1}{2},s)}(u)=\pm g^{(\frac{1}{2},s)}(-u-\eta)$, we conclude that the
eigenvalues of $\tilde t^{(\frac{1}{2},s)}(u)$ are
given by	
\be
\tilde \Lambda^{(\frac{1}{2},s)}(u) = 
\tilde h(u) {Q(u+p\eta)\over Q(u)} +
\tilde h(-u-\eta) {Q(u -p\eta)\over Q(u)} \,,
\label{TQ4}
\ee
where
\be 
\tilde h(u) &=&  \tilde h_{0}(u) h_{1}(u)
\ee
and
\be
\tilde h_{0}(u) = (-1)^{2sN} 4\sh^{2N}(u+(s+\frac{1}{2})\eta) 
{\sh(2u+2\eta)\over \sh(2u+\eta)}
\ee
The ansatz for $Q(u)$ is given by
\be
Q(u) = \prod_{j=1}^{M} 
\sh \left( {1\over 2}(u - u_{j}) \right)
\sh \left( {1\over 2}(u + u_{j} + \eta) \right) \,,
\label{Qbeta}
\ee 
which satisfies $Q(u + 2i\pi) = Q(u)$ and $Q(-u-\eta) = Q(u)$; and
\be
M=2sN+p
\,. \label{Mbeta}
\ee
Moreover, the Bethe ansatz equations for the zeros $u_{j}$ take the form 
\be
{\tilde h(u_{j})\over \tilde h(-u_{j}-\eta)} = 
-{Q(u_{j}-p\eta)\over Q(u_{j}+p\eta)} \,, 
\qquad j = 1 \,, \ldots \,, M \,.
\label{BAeqsbeta}
\ee
where we find the number of Bethe roots (\ref{Mbeta}) is consistent with the asymptotic behaviour (\ref{asymptotic}).

\subsection{One arbitrary $\beta$ and one arbitrary $\alpha$}\label{subsec:alphabetapm}

Finally, we consider combinations where the arbitrary parameters consist of one of the $\beta$'s and any one of  the $\alpha$'s. To keep the expressions general, we drop the subscripts $\pm$ from the boundary parameters, ${\alpha_{\pm}, \beta_{\pm}}$. The remaining boundary parameters are fixed, e.g., {${\beta_{+},\alpha_{-}}$} arbitrary, $\beta_{-} = \eta$, $\alpha_{+} = {i\pi\over 2}$ or other similar combinations. Also, as in previous cases, we let $\theta_{-} = \theta_{+} = \theta$. The matrix $\cal M$ is identical in form as in (\ref{calMalpha}). 

We once again find $h(u) = h_{0}(u)h_{1}(u)$, with the same $h_{0}(u)$ as for the earlier cases. For $h_{1}(u)$, we take the following,
\be
h_1(u) = \left\{ 
\begin{array}{ll}
    \ch u\ch(u-\eta) (\sh u+(-1)^{N}i\ch\beta)\sh(u-\alpha){\ch\left({1\over 2}(u+\alpha+\eta) \right)\over 
 \ch\left({1\over 2}(u-\alpha-\eta) \right)}\,, \\
    \qquad \qquad s= {1\over 2}\,, {3\over 2}\,, {5\over 2}\,, \ldots \\
   \ch u\ch(u-\eta) (\sh u+i\ch\beta)\sh(u-\alpha){\ch\left({1\over 2}(u+\alpha+\eta) \right)\over 
 \ch\left({1\over 2}(u-\alpha-\eta) \right)}\,, \\
\qquad \qquad s= 1\,, 2\,, 3\,, \ldots \\
\end{array} \right.
\label{hbetaalpha}
\ee
The above $h(u)$ satisfies (\ref{cond0})-(\ref{cond2}). The eigenvalues of the transfer matrix and Bethe ansatz equations are given by (\ref{eigenvalues}), (\ref{TQ3}), (\ref{Q}) and (\ref{BAeqs}), with 
\be
M = 2Ns+p
\label{Malfabeta}
\ee    
which again is consistent with (\ref{asymptotic}). We note that for $s = 1/2$, our solutions for all the above cases coincide with the corresponding solutions found in \cite{MNS}.

\section{Energy eigenvalues and Bethe roots}

In this section, we illustrate the completeness of the Bethe ansatz solutions derived in Sec. 3. We provide numerical evidence for cases $s = 1/2$ and $s = 1$, namely the complete energy levels together with the Bethe roots used in the computation (see Tables 1 and 2).

\subsection{s = 1/2 case}

The Hamiltonian for the open spin-$1/2$ XXZ quantum spin chain is given by \cite{dVGR, GZ}

\be
{\cal H} &=& {1\over 2}\sum_{n=1}^{N-1}\left( 
\sigma_{n}^{x}\sigma_{n+1}^{x}+\sigma_{n}^{y}\sigma_{n+1}^{y} 
+\ch \eta\ \sigma_{n}^{z}\sigma_{n+1}^{z}\right) \non \\
&+& {1\over 2}\sh \eta \Big[ 
\cth \alpha_{-} \tnh \beta_{-}\sigma_{1}^{z}
+ \csch \alpha_{-} \sech \beta_{-}\big( 
\ch \theta_{-}\sigma_{1}^{x} 
+ i\sh \theta_{-}\sigma_{1}^{y} \big) \non \\
& & \quad -\cth \alpha_{+} \tnh \beta_{+} \sigma_{N}^{z}
+ \csch \alpha_{+} \sech \beta_{+}\big( 
\ch \theta_{+}\sigma_{N}^{x}
+ i\sh \theta_{+}\sigma_{N}^{y} \big)
\Big]  \,, \label{Hamiltonianshalf} 
\ee
where $\sigma^{x} \,, \sigma^{y} \,, \sigma^{z}$ are the
standard Pauli matrices, $\eta$ is the bulk anisotropy parameter,
$\alpha_{\pm} \,, \beta_{\pm} \,, \theta_{\pm}$ are arbitrary boundary
parameters, and $N$ is the number of spins.

We compute the energy eigenvalues of (\ref{Hamiltonianshalf}) (from Bethe ansatz) for a particular case derived in Sec. 3. For the purpose of illustration, it is sufficient to consider the case where the two arbitrary boundary parameters are $\alpha_{-}$ and $\beta_{-}$. The steps here can be repeated for any other desired combinations of boundary parameters. The Hamiltonian (\ref{Hamiltonianshalf}) is related to the first derivative of the transfer matrix, $\tilde t^{(\frac{1}{2},\frac{1}{2})}(u)\cite{Sk},$\footnote{Note that for $s=1/2$, $t^{(\frac{1}{2},\frac{1}{2})}(u)=\tilde t^{(\frac{1}{2},\frac{1}{2})}(u)$}
\be
{\cal H} = c^{(\frac {1}{2})}_{1} {d \over du} \tilde t^{(\frac{1}{2},\frac{1}{2})}(u) \Big\vert_{u=0} 
+ c^{(\frac {1}{2})}_{2} \id \,,
\label{firstderivative}
\ee
where
\be
c^{(\frac {1}{2})}_{1} &=& -{1\over 16 \sh \alpha_{-} \ch \beta_{-}
\sh \alpha_{+} \ch \beta_{+} \sh^{2N-1} \eta 
\ch \eta} \,, \non \\
c^{(\frac {1}{2})}_{2} &=& - {\sh^{2}\eta  + N \ch^{2}\eta\over 2 \ch \eta} 
\,,
\label{cees}
\ee 
and $\id$ is the identity matrix. Moreover, (\ref{firstderivative}) implies that the energy eigenvalues are given by
\be
E = c^{(\frac {1}{2})}_{1} {d \over du} \tilde \Lambda^{(\frac{1}{2},\frac{1}{2})}(u) \Big\vert_{u=0} 
+ c^{(\frac {1}{2})}_{2} \,,
\label{firstderivative2}
\ee
Hence, using the results (\ref{TQ3})-(\ref{Q}) and (\ref{hbetaalpha})\footnote{The function $\tilde h(u)$ used here coincides with the one found in \cite{MNS}.} one arrives at the following result for the energy eigenvalues in terms of Bethe roots $\{u_{j}\}$,
\be
E &=& {1\over 2} \sh\eta\ch{\eta\over 2} \sum_{j=1}^{M}{1\over 
\sh ({1\over 2} u_{j} )\ch ({1\over 2} (u_{j} + \eta) )}
 + {1\over 2}N \ch \eta -{1\over 2}{\ch 2\eta\over \ch \eta} \non \\
&-& {1\over 2}\sh \eta(\coth \alpha_{-} + i\sech \beta_{-} - \tanh ({\alpha_{-}+\eta\over 2}))\,.
\label{energyspinhalf}
\ee
where $M = N + p$ (see (\ref{Malfabeta})).

In Table 1, we tabulate the energy eigenvalues computed using (\ref{energyspinhalf}) for $N = 4$ together with the Bethe roots (These roots are obtained using a method developed by McCoy and his collaborators \cite{McCoy} which is also explained in \cite{NR}.). This numerical result illustrates the completeness of Bethe ansatz equations derived in Sec. 3. We have verified that the energies given in Table 1 coincide with those obtained from direct diagonalization of (\ref{Hamiltonianshalf}). 

\subsection{s = 1 case}

In this section, we repeat the analysis for $s = 1$. We shall consider the case investigated in Sec. 3.2, namely the case with arbitrary $\beta_{-}$, $\beta_{+}$. The integrable Hamiltonian for the open spin-1 XXZ quantum spin chain is given by (adopting notations used in \cite{FNR})
\be
{\cal H} = \sum_{n=1}^{N-1}H_{n,n+1} + H_{b} \,.
\label{Hamiltonianspin1}
\ee
$H_{n,n+1}$ represents the bulk terms. Explicitly, these terms are given by \cite{ZF},
\be 
H_{n,n+1} &=&  \sigma_{n} - (\sigma_{n})^{2}
+ 2 \sh^2 \eta \left[ \sigma_{n}^{z} + (S^z_n)^2
+ (S^z_{n+1})^2 - (\sigma_{n}^{z})^2 \right] \non \\
&-& 4 \sh^2 (\frac{\eta}{2})  \left( \sigma_{n}^{\bot} \sigma_{n}^{z}
+ \sigma_{n}^{z} \sigma_{n}^{\bot} \right) \,, \label{bulkhamiltonianspin1}
\ee 
where
\be
\sigma_{n} = \vec S_n \cdot \vec S_{n+1} \,, \quad
\sigma_{n}^{\bot} = S^x_n S^x_{n+1} + S^y_n S^y_{n+1}  \,, \quad
\sigma_{n}^{z} = S^z_n S^z_{n+1} \,, 
\ee 
and $\vec S$ are the $su(2)$ spin-1 generators. $H_{b}$ represents the boundary terms which have the following form (see e.g., \cite{FNR, IOZ})
\be 
H_{b} &=& a_{1} (S^{z}_{1})^{2}  + a_{2} S^{z}_{1} 
+  a_{3} (S^{+}_{1})^{2}  +  a_{4} (S^{-}_{1})^{2}  +
a_{5} S^{+}_{1}\, S^{z}_{1}  + a_{6}  S^{z}_{1}\, S^{-}_{1} \non \\
&+& a_{7}  S^{z}_{1}\, S^{+}_{1} + a_{8} S^{-}_{1}\, S^{z}_{1} 
+ (a_{j} \leftrightarrow b_{j} \mbox{ and } 1 \leftrightarrow N) \,,
\ee
where $S^{\pm} = S^{x} \pm i S^{y}$. The coefficients $\{ a_{i} \}$ 
of the boundary terms at site 1 are functions  
of the boundary parameters ($\alpha_{-}, \beta_{-},
\theta_{-}$) and the bulk anisotropy parameter $\eta$. They are given by,
\be
a_{1} &=& \frac{1}{4} a_{0} \left(\ch 2\alpha_{-} - \ch 
2\beta_{-}+\ch \eta \right) \sh 2\eta 
\sh \eta \,,\non \\
a_{2} &=& \frac{1}{4} a_{0} \sh 2\alpha_{-} \sh 2\beta_{-} \sh 2\eta \,, \non \\
a_{3} &=& -\frac{1}{8} a_{0} e^{2\theta_{-}} \sh 2\eta 
\sh \eta \,, \non \\
a_{4} &=& -\frac{1}{8} a_{0} e^{-2\theta_{-}} \sh 2\eta 
\sh \eta \,, \non \\
a_{5} &=&  a_{0} e^{\theta_{-}} \left(
\ch \beta_{-}\sh \alpha_{-} \ch {\eta\over 2} +
\ch \alpha_{-}\sh \beta_{-} \sh {\eta\over 2} \right)
\sh \eta \ch^{\frac{3}{2}}\eta \,, \non \\
a_{6} &=&  a_{0} e^{-\theta_{-}} \left(
\ch \beta_{-}\sh \alpha_{-} \ch {\eta\over 2} +
\ch \alpha_{-}\sh \beta_{-} \sh {\eta\over 2} \right)
\sh \eta \ch^{\frac{3}{2}}\eta \,, \non \\
a_{7} &=&  -a_{0} e^{\theta_{-}} \left(
\ch \beta_{-}\sh \alpha_{-} \ch {\eta\over 2} -
\ch \alpha_{-}\sh \beta_{-} \sh {\eta\over 2} \right)
\sh \eta \ch^{\frac{3}{2}}\eta \,, \non \\
a_{8} &=&  -a_{0} e^{-\theta_{-}} \left(
\ch \beta_{-}\sh \alpha_{-} \ch {\eta\over 2} -
\ch \alpha_{-}\sh \beta_{-} \sh {\eta\over 2} \right)
\sh \eta \ch^{\frac{3}{2}}\eta \,,
\ee
where 
\be
a_{0}= \left[
\sh(\alpha_{-}-{\eta\over 2})\sh(\alpha_{-}+{\eta\over 2})
\ch(\beta_{-}-{\eta\over 2})\ch(\beta_{-}+{\eta\over 2})\right]^{-1} 
\,.
\ee
Similarly, the coefficients $\{ b_{i} \}$ of the boundary terms at 
site $N$ which are functions of  
the boundary parameters ($\alpha_{+}, \beta_{+}, \theta_{+}$) and $\eta$, are given by the following correspondence,
\be
b_{i} = a_{i}\Big\vert_{\alpha_{-}\rightarrow \alpha_{+}, 
\beta_{-}\rightarrow -\beta_{+}, \theta_{-}\rightarrow \theta_{+}} \,.
\ee

To derive the energy formula similar to (\ref{energyspinhalf}) for $s = 1$ case, we once again  begin by expressing the spin-1 Hamiltonian in terms of the first derivative of spin -1 transfer matrix, namely $t^{(1,1)}(u)$. One can construct $t^{(1,1)}(u)$ from $t^{(\frac {1}{2},1)}(u)$ by using the fusion hierarchy formula (\ref{hierarchy}),

\be
t^{(1,1)}(u) = t^{(\frac {1}{2},1)}(u-{\eta\over 2})t^{(\frac {1}{2},1)}(u+{\eta\over 2}) - \delta^{(1)}(u-{\eta\over 2})
\label{fh}
\ee
where
$\delta^{(1)}(u)$ is given by (\ref{dd})-(\ref{delta01}) with $s = 1$. Following \cite{FNR}, we work with rescaled transfer matrix given by
\be
\tilde t^{(1,1)\ gt}(u) = {\sh(2u) \sh(2u+2\eta)\over [\sh u 
\sh(u+\eta)]^{2N}}\, 
t^{(1,1)\ gt}(u) \,,
\label{rescaled}
\ee
where $t^{(1,1)\ gt}(u)$ is the transfer matrix constructed from ``gauge''-transformed $R^{(1,1)}(u)$ and $K^{\mp(1)}(u)$ matrices \footnote{One reason for such a transformation is to bring these matrices to a more symmetric form. For a detailed discussion on this, refer to Sec. 4 of \cite{FNR}}. Note: The rescaled transfer matrix does not vanish at $u=0$.
The Hamiltonian ${\cal H}$ (\ref{Hamiltonianspin1}), according to \cite{Sk}, is related to the first derivative of $\tilde t^{(1,1)\ gt}(u)$,
\be
{\cal H} = c^{(1)}_{1} {d \over du} \tilde t^{(1,1)\ gt}(u) \Big\vert_{u=0} 
+ c^{(1)}_{2} \id \,,
\label{firstderivatives1}
\ee
which in turn implies that the energy eigenvalues in terms of transfer matrix eigenvalues $\tilde \Lambda^{(1,1)\ gt}(u)$, are given by
\be
E = c^{(1)}_{1} {d \over du} \tilde \Lambda^{(1,1)\ gt}(u) \Big\vert_{u=0} 
+ c^{(1)}_{2}\,,
\label{firsteigenvaluess1}
\ee
where 
\be 
c^{(1)}_{1}&=&\ch \eta \Big\{ 16 [\sh 2\eta \sh \eta]^{2N} \sh 3\eta 
\sh(\alpha_{-}-{\eta\over 2})\sh(\alpha_{-}+{\eta\over 2})
\ch(\beta_{-}-{\eta\over 2})\ch(\beta_{-}+{\eta\over 2})\non \\
&\times& \sh(\alpha_{+}-{\eta\over 2})\sh(\alpha_{+}+{\eta\over 2})
\ch(\beta_{+}-{\eta\over 2})\ch(\beta_{+}+{\eta\over 2})\Big\}^{-1}
\label{c1sone}
\,.
\ee
and
\be 
c^{(1)}_{2}&=& -{a_{0}\over 4}b\ch\eta - (N-1)(4+\ch 2\eta) + 2 N \ch^{2}\eta \non \\
&-& {\sh\eta\over 2d}\Big\{-2\ch 2\alpha_{+}\Big(\ch\eta (3+7\ch 2\eta +\ch 4\eta)+\ch 2\beta_{+}(4+5\ch 2\eta+2\ch 4\eta)\Big)\non \\
&+& 2\ch \eta\Big(\ch 2\beta_{+}(3+7\ch 2\eta +\ch 4\eta)+\ch\eta (5+3\ch 2\eta +3\ch 4\eta)\Big)\Big\}\non \\
&-& {\sh 2\eta\over 2d}\Big\{\ch 2\beta_{+}(2+4\ch \eta \ch 3\eta)+\ch \eta (5\ch 2\eta +\ch 4\eta)-2\ch 2\alpha_{+}\Big(1+\ch 2\eta \non \\
&+& \ch 2\beta_{+}(\ch \eta +2\ch 3\eta)+\ch 4\eta\Big)\Big\}
\,.
\ee
where
\be
b = 2\big(-\ch 2\beta_{-}-\ch^{3}\eta + \ch 2\alpha_{-}(1+\ch 2\beta_{-}\ch\eta)\big)  
\ee
and
\be
d = -4\sh 3\eta \sh(\alpha_{+}+{\eta\over 2})\sh(\alpha_{+}-{\eta\over 2})
\ch(\beta_{+}+{\eta\over 2})\ch(\beta_{+}-{\eta\over 2})
\ee
Furthermore, using the fact that $\Lambda^{(1,1)\ gt}(u) = \Lambda^{(1,1)}(u)$ and (\ref{TQ4}), (\ref{fh}), (\ref{rescaled}), we obtain the energy in terms of Bethe roots $\{u_{j}\}$,
\be
E &=& -{1\over 2} \sh\eta\sh 2\eta \sum_{j=1}^{M}{1\over 
\ch ({1\over 2} (u_{j}+{3\eta\over 2}))\ch ({1\over 2} (u_{j} - {\eta\over 2}) )}
 + {1\over 2}{\sh 2\eta\over \tilde {\tilde h}({\eta\over 2})\tilde {\tilde h}(-{\eta\over 2})}(A'(0)+B'(0))\non \\
&+& c^{(1)}_{1}C'(0)+c^{(1)}_{2}\,.
\label{energyspinone}
\ee
where 
\be
A(u) &=& \tilde{\tilde h}(u+{\eta\over 2})\tilde{\tilde h}(u-{\eta\over 2})\non \\
B(u) &=& -\tilde{\tilde h}(u+{\eta\over 2})\tilde{\tilde h}(-u-{\eta\over 2})\non \\
C(u) &=& -\Big({\ch(u+{\eta\over 2})-i\sh\beta_{-}\over \ch(u+{\eta\over 2})+i\sh\beta_{-}}\Big)\Big({\ch(u+{\eta\over 2})+i\sh\beta_{+}\over \ch(u+{\eta\over 2})-i\sh\beta_{+}}\Big)B(u)\non \\
\tilde {\tilde h}(u) &=& 4\sh^{2N}(u+{3\eta\over 2})\sh(2u+2\eta)\sh(u+\eta)\sh(u-\eta)\non \\
&\times & (\ch u+i\sh\beta_{-})(\ch u-i\sh\beta_{+})
\ee
Also, $M = 2N + p$ (see (\ref{Mbeta})).

We tabulate the energies computed using (\ref{energyspinone}) for $N = 3$ with the Bethe roots (which are obtained using similar method as for the $s=1/2$ case above) in Table 2. These numerical results once again illustrate the completeness of Bethe ansatz equations derived in Sec. 3. We have verified that the energies given in Table 2 coincide with those obtained from direct diagonalization of (\ref{Hamiltonianspin1}). One can proceed to repeat the analysis for higher spin values, namely $s>1$. However, due to tedious computations, we avoid from pursuing it here.

\section{Discussion}\label{sec:discuss}

We have determined Bethe ansatz solutions of the open spin-$s$ XXZ quantum spin chain for cases with nondiagonal boundary terms (\ref{TQ3})-(\ref{BAeqs}) and (\ref{TQ4})-(\ref{BAeqsbeta}), by following the method used earlier in \cite{Ne, MNS} to solve the spin-$1/2$ case. This method relies on functional relations (\ref{funcrltn}) that the ``fundamental'' transfer matrices, $t^{(\frac{1}{2},s)}$(u) obey at roots of unity. However, these solutions hold only for $\eta = {i \pi\over 3}\,, {i \pi\over 5}\,,\ldots $. Unlike Bethe ansatz solutions found in earlier works on the open spin-$s$ XXZ chain with nondiagonal boundary terms, we emphasize that Bethe ansatz solutions found here hold for arbitrary values of boundary parameters (atmost two). We have checked these solutions for chains of length up to $N = 4$, and have verified that indeed they give the complete set of $(2s+1)^{N}$ eigenvalues. Moreover, we also presented numerical evidence for the completeness of the Bethe ansatz solutions found (using $s = 1/2$ and $s = 1$ as examples) in Tables 1 and 2.  Perhaps the completeness of the Bethe ansatz equations for spin-$s$ can readily be concluded from completeness of the corresponding Bethe ansatz equations for spin-$1/2$ case and the fusion hierarchy (\ref{hierarchy}) which is used in the construction of higher spin-$s$ transfer matrices.

There remain many problems worth investigating. As mentioned in the Introduction, due to the relation of $s = 1$ case to supersymmetric sine-Gordon (SSG) model, one can carry out similar analysis as in \cite{ANS}, but now for spin-$1$ chain with nondiagonal boundary terms. One could also try to extend the solutions presented here to cases with multiple $Q(u)$'s as in \cite{MNS2}. Also, to our knowledge, conventional form of Bethe ansatz solution for the open XXZ quantum spin chain, where all six boundary parameters are arbitrary with generic values of bulk anisotropy parameter $\eta$, has not been found. We remark that through a series of important work on the spectrum of XXZ spin chain based on representation theory of the q-Onsager algebra \cite{BK}, Baseilhac and Koizumi argue that obtaining such a conventional Bethe ansatz solution for the most general case is unlikely. It would be interesting to explore their results further and compare their approach with the Bethe ansatz approach.      

\section*{Acknowledgments}
I would like to thank R. I. Nepomechie for useful suggestions. I also thank P. Baseilhac for crucial correspondence.

\newpage
\begin{table}[htb] 
	    \centering
	    \begin{tabular}{|c|c|}\hline
	     $E$ &  Bethe roots ${u}_{j}$\\
	      \hline
	      -3.19769 & 0.222018 + 2.91719 i, 0.0900395 + 2.91719 i, - 2.6018 i,\\ & 1.01834 - 1.7952 i, 2.15279 i, 
                         0.267003 - 1.7952 i,\\ & 1.00769 i, 1.0165 + 1.3464 i, 0.0900395 - 0.224399 i,\\ & 0.222018-                         0.224394 i\\  
	     -2.42188 & 0.324807 - 3.13487 i, 0.319576 + 2.68662 i, 0.0958764 + 2.91717 i,\\ &
                       -2.60637 , 2.15279 i, 0.356519 - 1.7952 i,\\ & 1.09238 i, 0.0958764 - 0.224378 i, 0.324807-0.455523                         i,\\ & 0.319576 + 0.00617674 i\\
	      -1.87006 & 0.530712 + 2.91722 i, 0.0853747 + 2.91719 i, -2.60166 i,\\ & 0.946517 - 1.7952 i, 2.15279 i, 0.25634              - 1.7952 i,\\ & 1.00255 i, 0.943646 + 1.3464 i, 0.0853747 - 0.224399 i,\\ & 0.530712 - 0.224428 i\\
	      -1.30053 & 0.0805934 + 2.91719 i, 1.4454 - 1.7952 i, -2.6016 i,\\ & 0.607877 - 1.7952 i, 0.212672 - 1.7952 i, 
                         0.592602 + 1.3464 i,\\ & 0.992082 i, 0.54 i, 1.44531 + 1.3464 i,\\ & 0.0805934 - 0.224399 i \\
	      -0.874711 & 0.284541 + 2.91781 i, 0.251638 - 3.13998 i, 0.245572 + 2.69145 i,\\ & 2.15279 i, 0.359151 - 1.7952 i, 1.59561 i,\\ & -0.983447 i, 0.251638 - 0.450414 i, 0.245572 + 0.00134169 i,\\ & 0.284541 - 0.225016 i \\
	      -0.674656 & 0.518223 + 2.91722 i, 0.199566 + 2.91719 i, 0.940199 - 1.7952 i,\\ & 2.15279 i, 0.255178 - 1.7952 i, 1.00201 i,\\ & -0.988749 i, 0.937209 + 1.3464 i, 0.199566 - 0.2244 i, \\ & 0.518223 - 0.224431 i\\
	      -0.203476 & 0.182373 + 2.91719 i, 1.43967 - 1.7952 i, 0.604391 - 1.7952 i,\\ & 0.211794 - 1.7952 i, 0.588802 + 1.3464 i, 0.991943 i, -0.988797 i,\\ & 0.54 i, 1.43958 + 1.3464 i, 0.182373 - 0.224399 i\\
	
\hline
		  \end{tabular}
		  \caption[xxx]{\parbox[t]{0.8\textwidth}{
		  The 16 energies and corresponding Bethe roots 
		  given by $\tilde \Lambda^{(\frac{1}{2},\frac{1}{2})}(u)$ for  
		  $N=4\,, s=1/2\,, p=6\,, \eta = i\pi/7\,,
		  \alpha_{-}=0.54i\,, \beta_{-}=0.2\,, 
		  \theta_{-}=0\,, \alpha_{+}=i\pi/2\,, 
		  \beta_{+}=\eta\,, \theta_{+}=0 
		  $}
		  }
		 \label{table:energiesM}
\end{table}  
  \begin{table}[htb] 
	    \centering
	    \begin{tabular}{|c|c|}\hline
	     $E$ (continued) &  Bethe roots ${u}_{j}$ (continued)\\
	      \hline

                         0.343441 & 0.149446 - 3.14159 i, 0.149313 + 2.69279 i, 1.05439 - 1.7952 i,\\ & 2.15279 i, 0.273366 - 1.7952 i,
1.01147 i,\\ & 1.05294 + 1.3464 i, -2.60198 i, 0.149446 - 0.448806 i, \\ & 0.149313 \\
	      0.761262 & 0.0287807 - 2.72544 i, 0.487517 - 1.7952 i, 0.290846 + 1.3464 i,\\ & 0.0287807 - 0.864949 i, -0.619831 i, 0.54 i,\\ & 0.278493 i, -0.277755 i, -0.0641203 i, 0.0641203 i \\
	      0.846541 & 0.249771 - 3.14157 i, -3.01053 i, 3.01052 i,\\ & 0.186824 - 2.39256 i, 0.259537 + 2.25 i, 2.3744 i,\\ & 0.186824 - 1.19783 i, 0.54 i, 0.249771 - 0.44882 i,\\ & 0.259537 + 0.442796 i\\
	      0.883622 & 0.357565 + 3.13829 i, 0.357656 + 2.69611 i, 0.985138 - 1.7952 i,\\ &
2.15279 i, 0.260208 - 1.7952 i, 1.00394 i,\\ & -0.98875 i, 0.982836 + 1.3464 i, 0.357565 - 0.445494 i,\\ &
              0.357656 - 0.00331156 i\\
	      1.00689 &  0.351296 + 2.91719 i, 1.42001 - 1.7952 i, 0.593281 - 1.7952 i,\\ &
0.209089 - 1.7952 i, 0.576676 + 1.3464 i, 0.991553 i,\\ & -0.988798 i, 0.54 i, 1.41991 + 1.3464 i,\\ & 0.351296 - 0.224399 i\\
	      1.20648 & 0.452936 + 2.92003 i, 0.43968 - 2.9168 i, 0.439123 + 2.46302 i,\\ & - 2.6019 i, 2.15279 i, 0.328707 - 1.7952 i,\\ &  1.04938 i, 0.43968 - 0.673593 i, 0.439123 + 0.229771 i,\\ & 0.452936 - 0.227233 i \\
	      1.50502 & 0.625154 - 3.1108 i, 0.624995 + 2.66205 i, -2.6016 i,\\ & 0.830961 - 1.7952 i, 0.238271 - 1.7952 i, 1.69665 i,\\ & 0.825207 + 1.3464 i, 0.54 i, 0.625154 - 0.479587 i,\\ & 0.624995 + 0.0307393 i\\
	      1.82374 &  0.755163 + 2.9172 i, -2.60159 i, 1.3168 - 1.7952 i,\\ & 0.550839 - 1.7952 i, 0.199615 - 
1.7952 i, 0.530191 + 1.3464 i,\\ & 0.990548 i, 1.31659 + 1.3464 i, 0.54 i,\\ & 0.755163 - 0.224402 i\\

	      2.16601 & 1.72415 - 1.7952 i, 0.893009 - 1.7952 i, 0.447923 - 1.7952 i,\\ & 0.176866 - 1.7952 i, 1.70344 i,
0.416589 + 1.3464 i,\\ & 0.891068 + 1.3464 i, -0.988799 i, 0.54 i,\\ & 1.72413 + 1.3464 i\\
                  \hline
		  \end{tabular}
                  \end {table}  
\begin{table}[htb] 
	    \centering
	    \begin{tabular}{|c|c|}\hline
	     $E$ &  Bethe roots ${u}_{j}$\\
	      \hline
	      -12.4557 & 0.484779 - 3.10162 i, 0.411886 + 2.48641 i, 0.106801 - 3.14045 i,\\ & 0.0868008 + 2.51253 i, 0.348418 - 1.88496 i, 0.640811 + 1.25664 i, \\ &
0.106801 - 0.629463 i, 0.0868008 + 0.000739893 i, 0.484779 - 0.668294 i,\\ &
   0.411886 + 0.0268653 i\\  
	     -9.695 & 0.575021 + 2.83026 i, 0.0719595
- 3.14151 i, 0.0613009 + 2.51321 i,\\ & 
0.80436 - 1.88496 i, 0.0365392
+ 1.72227 i, 0.0365392 + 0.791004 i, \\ &
0.943625 + 1.25664 i, 0.0719595 - 0.628405 i, 0.0613009 + 0.0000685 i,\\ & 0.575021- 0.316986 i\\
	      -8.36086 & 0.500526 + 2.8322 i, 0.0403235
- 2.64238 i, 0.0383788 + 2.64283 i, \\ & 
0.0383947 + 2.38377 i, 0.670337 - 1.88496 i, 0.0403235 - 1.12753 i, \\ &
0.873976 + 1.25664 i, 0.0383788 - 0.129554 i, 0.0383947 + 0.129508 i,\\ & 0.500526 - 0.318927 i\\
	      -7.49773 & 0.507933 + 2.83108 i, 0.196307
+ 3.14091 i, 0.207919 + 2.51414 i, \\ &
0.723661 - 1.88496 i, 0.0062148 + 1.85904 i, 0.899811 + 1.25664 i, \\ &
0.0062148 + 0.654233 i, 0.196307 - 0.627633 i, 0.207919 - 0.000868076 i,\\ &
   0.507933 - 0.317806 i \\
              -7.43354 & 1.31309 -1.88496 i, 0.0437839 -3.14159 i, 0.0387234 +2.51327 i,\\ & 0.46288 -1.88496 i, 0.00846623 +1.83461 i, 0.635229 +1.25664 i,\\ & 0.00846623 +0.678666 i, 0.0437839 -0.628325 i, 0.0387234, \\ & 1.33352 +1.25664 i\\  
	      -6.81246 & 1.28291 - 1.88496 i, 0.00540715
+ 2.56635 i, 0.00749169 - 2.55718 i, \\ &
0.0054066 + 2.46021 i, 0.418583 - 1.88496 i, 0.00749169 - 1.21273 i, \\ &
0.603838 + 1.25664 i, 0.0054066 + 0.0530653 i, 0.00540715 - 0.053073 i,\\ & 1.30642 + 1.25664 i \\
	      -6.75338 & 0.534826 - 3.13136 i, 0.366787 + 2.83126 i, 0.11739 - 2.52032 i, \\ &
0.0961446 + 2.51326 i, 0.433162 + 1.84874 i, 0.11739 - 1.24959 i, \\ &
0.0961446 + 0.0000111021 i, 
  0.433162 + 0.664539 i, 0.366787 - 0.317989 i,\\ & 0.534826 - 0.638555 i\\
	    \hline
		  \end{tabular}
		  \caption[xxx]{\parbox[t]{0.8\textwidth}{
		  The 27 energies and corresponding Bethe roots 
		  given by $\tilde \Lambda^{(\frac{1}{2},1)}(u)$ for  
		  $N=3\,, s=1\,, p=4\,, \eta = i\pi/5\,,
		  \alpha_{-}=\eta\,, \beta_{-}=0.35\,, 
		  \theta_{-}=0.54\,, \alpha_{+}=\eta\,, 
		  \beta_{+}=0.76\,, \theta_{+}=0.54 
		  $}
		  }
		 \label{table:energiesM2}
\end{table}  
                  
\begin{table}[htb] 
	    \centering
	    \begin{tabular}{|c|c|}\hline
	     $E$ (continued) &  Bethe roots ${u}_{j}$ (continued)\\
	      \hline

                -6.37707 & 0.450862 - 3.1307 i, 0.3548 + 
2.83363 i, 0.203444 + 3.14135 i, \\ &
0.313088 + 1.93502 i, 0.10853 + 
1.81328 i, 0.10853 + 0.699993 i, \\ &
0.203444 - 0.62808 i, 0.313088
+ 0.578258 i, 0.450862 - 0.639212 i,\\ &
   0.3548 - 0.320357 i\\
	      -5.9431 & 0.15167 + 3.14151 i, 1.28717 - 
1.88496 i, 0.152799 + 2.51336 i, \\ &
0.431361 - 1.88496 i, 0.000660592 + 1.88095 i, 
  0.610477 + 1.25664 i,\\ & 0.000660592 + 0.632323 i, 0.15167 - 
0.628234 i, 1.3102 + 1.25664 i, \\ &
0.152799 - 0.0000868625 i \\
	      
                 -4.96948 & 0.245899 - 3.14157 i, 1.09088 - 
1.88496 i, 0.0674913 - 2.51035 i, \\ &
0.056464 + 2.51327 i, 0.236615
+ 1.88684 i, 0.0674913 - 1.25957 i, \\ &
1.14613 + 1.25664 i, 0.236615 + 
0.626432 i, 0.245899 - 0.628337 i, \\ &
0.056464\\

	      -4.75362 & 0.118717 + 3.04878 i, 0.118716
- 3.04877 i, 1.08872 - 1.88496 i, \\ &
0.120353 + 1.97697 i, 0.0725214
+ 1.8268 i, 0.0725214 + 0.686472 i, \\ &
0.118716 - 0.721137 i,
   0.118717 - 0.535507 i, 0.120353
 + 0.536301 i,\\ &
   1.14446 + 1.25664 i\\  

	     -4.29889 & 0.611823 - 3.10726 i, 0.585252
+ 2.48522 i, 0.283735 + 2.82749 i, \\ &
0.545972 - 1.88496 i, 
0.00037444 + 1.88259 i,
0.773596 + 1.25664 i, \\ &
0.00037444 + 0.630686 i, 
  0.283735 - 0.314213 i, 0.611823 - 0.662648 i,\\ & 0.585252 + 0.0280531 i\\

	      -4.09589 & 0.792496 + 2.51421 i, 0.79911 - 
2.5049 i, 0.0126395 + 2.51327 i, \\ &
0.0132493 - 2.51136 i, 
0.0132493 - 1.25855 i, 
0.0832428 + 1.25664 i,\\ & 0.79911 - 1.26501 i, 0.939738 + 1.25664 i, 
0.0126395,\\ & 0.792496 - 0.000938409 i \\

	      -4.05775 & 0.79092 +2.51422 i, 0.797571 -2.50483 i, 0.0134204 +3.14159 i,\\ & 0.0128183 +1.8868 i, 0.0564763 +1.25664 i, 0.797571 -1.26508 i,\\ & 0.938917 +1.25664 i, 0.0134204 -0.628319 i, 0.0128183 +0.62647 i, \\ & 0.79092 -0.000943414 i\\

	      -3.93649 & 0.886933 -3.14031 i, 0.0107824 +2.51327 i, 0.0112295 -2.51164 i,\\ & 0.876475 +1.88715 i, 0.725212 -1.88496 i, 0.0112295 -1.25828 i,\\ & 0.074371 +1.25664 i, 0.0107824, 0.876475 +0.626128 i,\\ & 0.886933 -0.629601 i \\

	      -3.90338 & 0.886042 -3.1403 i, 0.011339 +3.14159 i, 0.875518 +1.88714 i,\\ & 0.723621 -1.88496 i, 0.0108981 +1.88655 i, 0.0516604 +1.25664 i,\\ & 0.011339 -0.628319 i, 0.0108981 +0.626729 i, 0.875518 +0.626131 i,\\ & 0.886042 -0.629609 i\\
      
                  \hline
		  \end{tabular}
                  \end {table}  

\begin{table}[htb] 
	    \centering
	    \begin{tabular}{|c|c|}\hline
	     $E$ (continued) &  Bethe roots ${u}_{j}$ (continued)\\
	      \hline
               
 -3.5973 & 0.336944 -3.14041 i, 0.335901 +2.51212 i, 1.23958 -1.88496 i,\\ & 0.391865 -1.88496 i, 0.000094209 +1.88434 i, 0.568642 +1.25664 i,\\ & 0.000094209 +0.628934 i, 1.26813 +1.25664 i, 0.336944 -0.629498 i,\\ & 0.335901 +0.00115798 i\\

	      -2.69459 & 0.779303 +2.82839 i, 0.232072 +2.82743 i, 1.05217 -1.88496 i,\\ & 0.329315 -1.88496 i, 0.0000460866 +1.88465 i, 0.475658 +1.25664 i,\\ & 0.0000460866 +0.628622 i, 1.10962 +1.25664 i, 0.232072 -0.31416 i, \\ & 0.779303 -0.315113 i\\

                 -2.39712 & 0.503174 +2.79201 i, 0.504144 -2.78706 i, 0.506179 +2.23281 i, \\ & 0.745784 -1.88496 i, 0.000133753 +1.88409 i, 0.944351 +1.25664 i, \\ & 0.000133753 +0.629187 i, 0.504144 -0.982846 i, 0.503174 -0.278731 i,\\ & 0.506179 +0.280461 i\\

	      -1.91101 & 0.745717 -3.13841 i, 0.476924 -2.52358 i, 0.476385 +2.51311 i, \\ & 0.712375 +1.88751 i, 0.000108768 +1.88425 i, 0.476924 -1.24633 i,\\ &  0.000108768 +0.629027 i, 0.476385 +0.000165845 i, 0.712375 +0.625767 i, \\ & 0.745717 -0.631505 i\\  

	     -1.65102 & 0.591714 +3.13021 i, 0.594377 +2.52503 i, 1.08954 -1.88496 i,\\ & 0.319176 -1.88496 i, 0.0000183025 +1.88483 i, 0.467049 +1.25664 i,\\ & 0.0000183025 +0.62844 i, 1.14296 +1.25664 i, 0.591714 -0.616939 i,\\ & 0.594377 -0.0117591 i\\

	      -1.3681 &  1.52581 -1.88496 i, 0.205857 +2.82743 i, 0.710251 -1.88496 i, \\ & 0.25594 -1.88496 i,
1.8849 i, 0.349688 +1.25664 i, \\ & 0.813986 +1.25664 i, 0.628371 i, 0.205857 -0.314159 i,\\ & 1.53319 +1.25664 i\\

	      -1.24743 & 0.75169 +2.86018 i, 0.738919 -2.84659 i, 0.731307 +2.15882 i,\\ &  0.34774 -1.88496 i, 0.0000213862 +1.88481 i, 0.564622 +1.25664 i,\\ & 0.0000213862 +0.62846 i, 0.738919 -0.92332 i, 0.731307 +0.354451 i,\\ & 0.75169 -0.346911 i\\

	      -0.451049 & 0.496986 +2.82743 i, 1.4849 -1.88496 i, 0.679662 -1.88496 i, \\ & 0.240941 -1.88496 i, 1.88494 i, 0.326913 +1.25664 i, \\ & 0.789576 +1.25664 i, 0.628339 i, 1.49385 +1.25664 i, \\ & 0.496986 -0.314155 i \\

	      -0.278905 & 0.901463 -3.10659 i, 0.894513 +2.47987 i, 0.816494 -1.88496 i, \\ & 0.258214 -1.88496 i, 1.88493 i, 0.358314 +1.25664 i, \\ & 0.927674 +1.25664 i, 0.628344 i, 0.901463 -0.663324 i,\\ & 0.894513 +0.0334058 i\\

                  \hline
		  \end{tabular}
                  \end {table}  

\begin{table}[htb] 
	    \centering
	    \begin{tabular}{|c|c|}\hline
	     $E$ (continued) &  Bethe roots ${u}_{j}$ (continued)\\
	      \hline

	      0.72331 & 1.02774 +2.82771 i, 1.29908 -1.88496 i, 0.601685 -1.88496 i, \\ & 0.212716 -1.88496 i, 1.88495 i, 0.282851 +1.25664 i, \\ & 0.716771 +1.25664 i, 0.628324 i, 1.31854 +1.25664 i,\\ & 1.02774 -0.314435 i\\

	      1.69087 & 1.74631 -1.88496 i, 0.943277 -1.88496 i, 0.514308 -1.88496 i, \\ & 0.183505 -1.88496 i, 1.88495 i, 0.238533 +1.25664 i, \\ & 0.614563 +1.25664 i, 0.62832 i, 0.990458 +1.25664 i, \\ & 1.7488 +1.25664 i\\
\hline
		  \end{tabular}
                  \end {table}

\end{document}